\documentclass[11pt,a4paper]{article}
\usepackage{jheppubnohead}
\usepackage[utf8]{inputenc}
\usepackage{epsfig}
\usepackage{tabularx}
 \usepackage{amsmath,amsfonts,amssymb,dsfont,mathrsfs,graphicx, booktabs, siunitx, caption, subcaption, slashed}
\usepackage[capitalise, english]{cleveref}

\makeatletter
\providecommand*{\diff}%
	{\@ifnextchar^{\DIfF}{\DIfF^{}}}
\def\DIfF^#1{%
	\mathop{\mathrm{\mathstrut d}}%
		\nolimits^{#1}\gobblespace}
\def\gobblespace{%
	\futurelet\diffarg\opspace}
\def\opspace{%
	\let\DiffSpace\!%
	\ifx\diffarg(%
		\let\DiffSpace\relax
	\else
		\ifx\diffarg[%
			\let\DiffSpace\relax
		\else
  			\ifx\diffarg\{%
				\let\DiffSpace\relax
			\fi\fi\fi\DiffSpace}

\newcommand{\AddrZurich}{%
Physik-Institut, Universit\"at Z\"urich, CH-8057 Z\"urich, Switzerland
}

\notoc
\title{Multi-Scale 5D Models for Flavor Hierarchies and Anomalies}

\author[a]{Ben A. Stefanek} \emailAdd{bestef@physik.uzh.ch}
\affiliation[a]{\AddrZurich}

\abstract{So-called 4321 gauge models at the TeV scale with hierarchical couplings reminiscent of the Standard Model Yukawas offer a coherent combined explanation of the recent $B$-meson anomalies. In these proceedings, based on arXiv:2203.01952, we discuss how such models could arise from a multi-scale theory of flavor, based on a warped fifth dimension with three branes. This higher dimensional construction provides a natural description of flavor hierarchies, addresses the electroweak hierarchy problem, and allows for the Higgs to be identified as a pseudo-Nambu-Goldstone boson emerging from the same dynamics responsible for the breaking of 4321 gauge symmetry.}

\begin{document}

\maketitle
\newpage

\section{Introduction}

In recent years a series of anomalies hinting at beyond the Standard Model (SM) lepton flavor universality violation in $B$-meson decays have emerged, both in charged and neutral currents. Interestingly, these $B$ anomalies can be coherently explained by new flavor non-universal interactions at the TeV scale, with a coupling structure reminiscent of the SM Yukawas, strongly suggesting a connection between the new physics sector and the SM flavor puzzle. One of the most promising explanations are new flavor non-universal gauge interactions, best exemplified in the context of “4321" gauge models, where the SM gauge symmetry is extended by an $SU(4)$ factor under which only the third family is charged. These models provide third-family quark-lepton unification, a natural CKM structure, and a UV completion at the TeV scale for the Pati-Salam vector leptoquark (LQ) (typically denoted as $U_1$). The latter offers what is to date the most coherent combined explanation for the $B$-meson anomalies. Interestingly, since third-family quark-lepton unification occurs at the TeV scale in order to address the $B$ anomalies, it is plausible that the same new physics sector plays a role in the resolution of the electroweak (EW) hierarchy problem. Such non-universal 4321 models could simply be the low-energy limit of a multi-scale theory of flavor, where quarks and leptons of different families are unified at sequentially higher energy scales. A compelling possibility is to connect these models to a full theory of flavor and the EW hierarchy problem in terms of a warped extra dimension with three defects (or branes) where each SM family is quasi-localized~\cite{Fuentes-Martin:2022xnb}. This novel 5-dimensional (5D) construction unifies all families of quark and leptons in a flavor non-universal manner, provides a natural description of the SM fermion masses and mixings, and solves the large hierarchy problem à la Randall-Sundrum (RS)~\cite{Randall:1999ee}. In stark contrast to the usual RS flavor anarchy paradigm, this multi-brane construction provides a novel explanation of the flavor hierarchies in terms of a hierarchy of scales. It also provides unique new experimental signatures, both at low energy as well as in cosmological observables.
\section{4321 models}
\begin{table}[h]
\begin{center}
\scalebox{1}{\begin{tabular}{|c|c|c|c|c|}
\hline
Field & $SU(4)_h$ & $SU(3)_l$ & $SU(2)_L$ & $U(1)_{X}$ \\
\hline
\hline
$q^i_L$ & $\mathbf{1}$ & $\mathbf{3}$ & $\mathbf{2}$ & $1/6$ \\
$u^i_R$ & $\mathbf{1}$ & $\mathbf{3}$ & $\mathbf{1}$ & $2/3$  \\
$d^i_R$ & $\mathbf{1}$ & $\mathbf{3}$ & $\mathbf{1}$ & $-1/3$  \\
$\ell^i_L$ & $\mathbf{1}$ & $\mathbf{1}$ & $\mathbf{2}$ & $-1/2$ \\
$e^i_R$ & $\mathbf{1}$ & $\mathbf{1}$ & $\mathbf{1}$ & $-1$ \\ 
$\psi_L$ & $\mathbf{4}$ & $\mathbf{1}$ & $\mathbf{2}$ & $0$ \\ 
$\psi_R^{\pm}$ & $\mathbf{4}$ & $\mathbf{1}$ & $\mathbf{1}$ & $\pm1/2$ \\  
$\chi_{L,R}$ & $\mathbf{4}$ & $\mathbf{1}$ & $\mathbf{2}$ & 0  \\  
\hline
\hline
$H$ & $\mathbf{1}$ & $\mathbf{1}$ & $\mathbf{2}$ & 1/2  \\    
$\Omega_1$ & $\mathbf{\bar 4}$ & $\mathbf{1}$ & $\mathbf{1}$ & $-1/2$  \\ 
$\Omega_3$ & $\mathbf{\bar 4}$ & $\mathbf{3}$ & $\mathbf{1}$ & $1/6$  \\  
$\Omega_{15}$ & $\mathbf{15}$ & $\mathbf{1}$ & $\mathbf{1}$ & 0  \\ 
\hline
\end{tabular}}
\end{center}
\caption{ Field content of the 4321 model. 
}
\label{tab:fieldcontent}
\end{table}
We begin with a brief overview of 4321 models~\cite{DiLuzio:2017vat,Bordone:2017bld,Greljo:2018tuh,Fuentes-Martin:2020bnh}, as they provide a compelling UV completion for the $U_1$ vector leptoquark that is known to provide a good combined explanation of the $B$-anomalies~\cite{Alonso:2015sja,Calibbi:2015kma,Barbieri:2015yvd,Bhattacharya:2016mcc,Buttazzo:2017ixm,DiLuzio:2018zxy,Cornella:2019hct,Cornella:2021sby}. They are based on the the extended gauge group
\begin{equation}
\mathcal{G}_{4321} = SU(4)_h \times SU(3)_l \times SU(2)_L \times U(1)_X \,,
\end{equation}
where the SM color group is embedded as $SU(3)_c = [SU(3)_h \times SU(3)_l]_{\rm diag}$ and SM hypercharge as $U(1)_Y = [U(1)_h \times U(1)_X]_{\rm diag}$, where $SU(3)_h \times U(1)_h \subset SU(4)_h$. The 4321 gauge group is spontaneously broken to the SM at the TeV scale by the VEVs of the $\Omega_{1,3,15}$ fields, resulting in 15 massive gauge bosons transforming as $U_1 \sim (\mathbf{3},\mathbf{1},2/3)$, $Z' \sim (\mathbf{1},\mathbf{1},0)$, and $G' \sim (\mathbf{8},\mathbf{1},0)$ under the SM gauge group. The field $G'$ is a heavy gluon that we call the coloron. Defining the gauge fields of the 4321 group as $H_{\mu}^\alpha,  g_{\mu}^{\prime a}, W_{\mu}^i, X_{\mu}$ and gauge couplings $g_4,g_3,g_2,g_1$, the massive $U_1, Z', G'$ gauge bosons are
\begin{align}
U_{\mu}^{1,2,3} &= \frac{1}{\sqrt{2}} \left( H_{\mu}^{9,11,13} - i H_{\mu}^{10,12,14} \right) \,, \nonumber \\
Z'_{\mu} &= H_{\mu}^{15} \cos\theta_1  - X_{\mu}\sin\theta_1  \,, \nonumber \\
G_{\mu}^{\prime a} &= H_{\mu}^{a}\cos\theta_3  - g_{\mu}^{\prime a}  \sin\theta_3 \,, 
\end{align}
with $\tan^2 \theta_{3} = g_{3}^2/g_4^2$ and $\tan^2\theta_{1} = 2g_{1}^2/3g_4^2$. In terms of these mixing angles, the gauge boson masses are
\begin{align}
M_U^2 = \frac{g_{4}^2 f_U^2}{4} \,, \hspace{15mm} M_{Z',G'}^2 = \frac{M_U^2}{\cos^2\theta_{1,3}} \frac{f_{Z',G'}^2}{f_U^2} \,,
\end{align}
with $f_U^2 = v_1^2 + v_3^2 +4v_{15}^2/3$, $f_{Z'}^2 =3 v_1^2 /2+ v_3^2/2$, and $f_{G'}^2 = 2v_{3}^2$. 
In the limit $f_U = f_{Z'} = f_{G'}$ (corresponding to $v_1 = v_3$ and $v_{15} = 0$) there is an unbroken $SU(4)_D$ custodial symmetry which defines a $\rho$-parameter-like relation between the 4321 gauge boson masses, $M_U^2 / M_{Z',G'}^2 = \cos^2\theta_{1,3}$.  As promised, the linear combinations orthogonal to $G_{\mu}^{\prime a} $ and $Z'_{\mu}$ remain massless and correspond to the $SU(3)_c \times U(1)_Y$ part of the SM gauge group. 

As suggested by the labels $``h"$ and $``l"$ on the color $SU(4)_h \times SU(3)_l$ part of the 4321 group, it is attractive to charge the (heavy) third family SM fermions under the $SU(4)_h$ factor, while keeping the light families SM-like by charging them under $SU(3)_l$, as shown in Table~\ref{tab:fieldcontent}. This leads to quark-lepton unification à la Pati-Salam in the third family, while also realizing an accidental approximate $U(2)^5 \equiv U(2)_q \times U(2)_\ell \times U(2)_u \times U(2)_d \times U(2)_e$ flavor symmetry as a consequence of the gauge charge assignment. Furthermore, in the limit where $g_4 \gg g_{1,3}$, the new heavy vectors $U_1, Z', G'$ come mostly from $SU(4)_h$ and are therefore dominantly coupled to the third family, as required to address the $B$-anomalies while avoiding direct search bounds from colliders. Indeed, with this charge assignment one is only allowed to write Yukawa couplings for the 3rd family as well as amongst the light families (mixing between the 3rd and light families is forbidden by 4321 gauge symmetry). Heavy-light mixing is generated by the following Lagrangian
\begin{align}\label{eq:LYuk}
\begin{aligned}
-\mathcal{L} & \supset   \bar q_{L}\lambda_{q} \Omega_{3} \chi_{R}  +  \bar \ell_{L} \lambda_{\ell} \, \Omega_{1}   \chi_{R} + {y}_{+} \bar \chi_L   \tilde{H} \psi_R^{+} + y_{-} \bar \chi_L H \psi_R^{-} + M_{\chi} \bar \chi_L \chi_R + \rm{h.c.}  \,,
\end{aligned}
\end{align}
after the 4321 gauge symmetry is spontaneously broken. Note that because of the quantum numbers of the VL fermion, mixing is only generated between the light family LH doublets $q_L, \ell_L$ and $\chi_L$, corresponding to a leading breaking of the $U(2)^5$ symmetry only in the left-handed (LH) sector. This leads to a natural suppression of dangerous FCNC operators involving light family right-handed (RH) fields (such as scalar operators and dipoles), as the corresponding rotations are suppressed by the light family Yukawas which are the only source of $U(2)_u \times U(2)_d \times U(2)_e$ breaking.

\section{A multi-scale paradigm for flavor in warped extra dimensions}
Warped extra dimensions represent one of the most natural frameworks in which solutions to the EW hierarchy problem as well as the flavor puzzle can be consistently combined. In the usual Randall-Sundrum setup, large hierarchies between the scales of the ultraviolet (UV) brane and the infrared (IR) brane can be naturally stabilized due to a gravitational redshift factor arising from the warped (${\rm AdS_5}$) geometry. Parametrically, this results in an IR scale $\Lambda_{\rm IR} = k e^{-kL}$, where $k$ is the 5D curvature constant (and also the typical scale of the UV brane) and $L$ is the total length of the extra dimension. Exponentially large hierarchies are therefore generated for modest values of the volume factor $kL$, and such a setup can be used to solve the large electroweak hierarchy problem if the Higgs is localized in the IR. Similarly, the observed hierarchy of fermion masses and mixings can be explained if the heavy fermions (such as the top quark) are localized in the IR while the lighter fermions are localized in the UV. Yukawa couplings must be generated at the IR scale (where the Higgs is localized), resulting in hierarchical values due to the exponential behavior of the fermion zero mode profiles. This mechanism of obtaining effective 4D Yukawa couplings with exponential hierarchies even when the original 5D Yukawas exhibit no such structure is known as the flavor anarchy paradigm. However, the flavor anarchy setup is not entirely satisfactory since both the left- and right-handed fields must reach the IR in order to generate the Yukawa couplings, corresponding to a breaking of $U(2)^5$ in both the left- and right-handed sectors. In general, this leads to dangerous operators involving the light family RH fields being generated at the IR scale, such as the electron EDM or charged lepton flavor violating processes such as $\mu \rightarrow e \gamma$. In order to avoid the bounds on such processes, one is forced to take $\Lambda_{\rm IR} > 50-100$ TeV, see e.g. Ref.~\cite{Panico:2016ull}. This can be done, but the price to pay is a large fine tuning of the EW scale. 
\begin{figure}[t]
\centering
\vspace{-15mm}
\includegraphics[width=0.925\columnwidth]{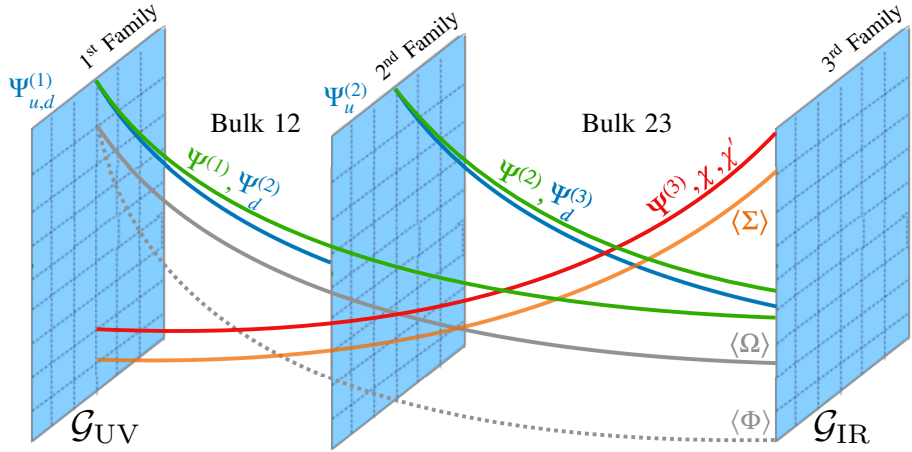}
\vspace{-18mm}
\caption{Sketch of a multi-scale warped 5D model including fermion and scalar profiles.}
\label{fig:branes}
\end{figure}

The situation can be improved by extending the Higgs profile into the bulk and moving to a multi-brane construction, with one brane per family, as shown in Fig.~\ref{fig:branes} (with the role of the Higgs played by $\Sigma$). In order from UV to IR, we now have three scales $\Lambda_1, \Lambda_2, \Lambda_{\rm IR}$. The major improvement comes from the fact that the RH fields can now be taken to be highly localized in their respective branes, resulting in the generation of the light family Yukawas at the UV scales $\Lambda_{1,2}$, with their smallness now explained by the exponentially falling Higgs profile. Since the Higgs profile falls according to the ratio of scales, this setup can be viewed equivalently as understanding the hierarchies in the Yukawa couplings as arising from a hierarchy of scales $y_{1,2} \sim \Lambda_{\rm IR} / \Lambda_{1,2}$. Interestingly, this fixes the total volume of the extra dimension $kL \approx \log(m_t / m_u) \approx 10 $ in order to provide the correct fermion mass hierarchy. Similarly, dangerous operators involving the light family RH fields such as the aforementioned dipoles must now be generated at (and are therefore suppressed by) the UV scales, allowing for a lower IR scale $\Lambda_{\rm IR}\sim$ TeV. Fermion mixing is explained via the quasi-localization of the LH fields in their respective branes, resulting in a nearest-neighbor-style mixing structure and corresponding to a breaking of $U(2)^5$ only in the LH sector (as in the 4D 4321 model). For fixed values of the LH fermion profiles, fermion mixing fixes the distances between branes, e.g. Cabibbo mixing fixes the distance between the 1st and 2nd family branes as $k\ell \approx 4$.

\section{A multi-scale warped 5D model}
We now want to construct a warped 5D model based on the following key features:
\begin{enumerate}
\item {\it 4321 gauge symmetry at the TeV scale.}
\item {\it Flavor hierarchies from a 3-brane structure in the fifth dimension.}
\item {\it SM Higgs as a pNGB from the same dynamics breaking 4321 gauge symmetry.}
\end{enumerate}
Since the benefits of the first two points have already been discussed above, we turn now to discussion regarding the third point. According to the duality between warped 5D theories and strongly-coupled 4D composite theories, a reduction of the 5D bulk gauge symmetry on the IR brane $\mathcal{G}_{{\rm bulk}} \rightarrow \mathcal{G}_{\rm IR}$ corresponds to a spontaneous breaking (SSB) of the global symmetry of the 4D composite sector  due to the formation of a condensate in the IR, resulting in Nambu-Goldstone bosons (NGBs) in the coset $\mathcal{G}_{{\rm bulk}}/ \mathcal{G}_{\rm IR}$. If a subgroup $\mathcal{G}_{\rm gauge}$ of the global symmetry is gauged, it is also possible for the condensate to break part of the gauge symmetry, giving mass to elementary gauge bosons à la technicolor models. The surviving unbroken gauge symmetry is $\mathcal{G}_{\rm 0} = \mathcal{G}_{\rm gauge} \cap \mathcal{G}_{\rm IR}$. In order to realize point 3, we would like to build a model where $\mathcal{G}_{\rm gauge} \supset \mathcal{G}_{\rm 4321}$, $\mathcal{G}_{\rm 0} = \mathcal{G}_{\rm SM}$, and the SM Higgs field is realized as pNGB living in the coset $\mathcal{G}_{{\rm bulk}} / \mathcal{G}_{\rm IR}$.
Since we have a gauge rather than global symmetry in the warped 5D description, realizing the Higgs a pNGB corresponds to gauge-Higgs unification~\cite{Contino:2003ve}, which can be achieved by extending the EW part of the bulk gauge symmetry. Since our multi-scale model actually has two bulks, this will be dual to a strongly-coupled theory that undergoes two SSBs, while $\mathcal{G}_{\rm gauge}$ will be the gauge symmetry in the most UV brane. A simple and minimal option is to consider 
\begin{align}\label{eq:GbulkIR}
\begin{aligned}
\mathcal{G}^{\rm 23}_{{\rm bulk}} &\equiv SU(4)_h\times SU(3)_l\times U(1)_l \times  SO(5)\,, \\
\mathcal{G}_{\rm IR} &\equiv SU(3)_c\times U(1)_{B-L} \times SO(4)\,,
\end{aligned}
\end{align} 
where the 23 bulk most IR bulk (see Fig.~\ref{fig:branes}). There are 15 NGBs coming from the breaking pattern $SU(4)_h\times SU(3)_l\times U(1)_l \rightarrow SU(3)_c\times U(1)_{B-L}$. These 15 NGBs are eaten by the 4321 gauge bosons $U_1, G', Z'$, which acquire masses of $M_{15} \approx M_{\rm KK} / \sqrt{2kL}$, with $M_{\rm KK} \approx 2\Lambda_{\rm IR}$, so there is a natural mass gap between the mass scale of the 4321 gauge bosons and that of the Kaluza-Klein (KK) resonances. A benchmark value providing a good explanation of the $B$-anomalies is $M_{15} \approx 3.5$ TeV, which results in $\Lambda_{\rm IR} \approx 8$ TeV and $M_{\rm KK} \approx 16$ TeV. The leading deviations from typical 4321 phenomenology come from EW KK-induced modification of the SM $Z$ boson coupling to $\tau_L$, which simply require $M_{\rm KK} \gtrsim 6$ TeV. Note that an explanation of the $B$-anomalies would be incompatible with the $Z\rightarrow \tau_L \tau_L$ bound if $M_{15} \approx M_{\rm KK}$, so models where the $U_1$ LQ is realized as a composite (KK) resonance face challenges with EW precision data. 

The breaking $SO(5) \rightarrow SO(4) \equiv SU(2)_L \times SU(2)_R$ was chosen in order to realize the minimal composite Higgs scenario~\cite{Agashe:2004rs}, where the fifth components of the gauge fields associated to the $SO(5)/SO(4)$ coset correspond to four pNGB zero modes transforming as a ${\bf 4}$ of $SO(4)$ that we identify with the SM Higgs field. The Higgs is a pNGB as it receives a potential at tree-level from $SO(5)$-breaking scalars, as well as at the loop level. These contributions are finite and fully calculable within the 5D model, and no double tuning is required- the Higgs quartic naturally comes out at the right size. Only the Higgs VEV must be fine-tuned at the per mille level, corresponding to the usual little hierarchy problem.

Moving deeper into the UV, it is attractive to choose
\begin{align}
\mathcal{G}^{\rm 12}_{{\rm bulk}} &\equiv SU(4)_h\times SU(4)_l\times SO(5) \,, \\  
\mathcal{G}_{\rm UV} & \equiv SU(4)_h\times SU(3)_l\times U(1)_l\times SU(2)_L\times U(1)_R\,, \nonumber
\end{align}
where $\mathcal{G}_{\rm UV}$ is the gauge symmetry in the most UV brane which contains $\mathcal{G}_{\rm 4321}$ as a subgroup. The middle brane in the chosen setup corresponds to a discontinuity inducing the breaking $SU(4)_l \rightarrow SU(3)_l \times U(1)_l$, or a phase transition where light-family quark-lepton unification is broken. This phase transition, as well as the breaking of third-family quark-lepton unification at the IR scale, could potentially be observable as a multi-peaked stochastic gravitational wave signal~\cite{Greljo:2019xan}.

\subsection{Matter Content}
The fermions and scalars of the model and their representations under $SU(4)_h\times SU(4)_l\times SO(5) $ are given in Table~\ref{tab:content}. The UV localized scalars $\Phi$ and $\Omega$ are necessary along with the singlet fermions $\mathcal{S}^i$ to give mass to neutrinos via an inverse-seesaw mechansism~\cite{Fuentes-Martin:2020pww}. The VEV of $\Omega$ also breaks $\mathcal{G}_{\rm UV}\rightarrow \mathcal{G}_{\rm 4321}$. The scalar $\Sigma$ is IR localized and plays a crucial role in generating the light family Yukawas, as we will see below. 
\begin{table}[t]
    \renewcommand{\arraystretch}{1.2}
    \centering
    \scalebox{1}{\begin{tabular}{|c|ccc||cc|}
    \hline
    Field & $SU(4)_h$ & $SU(4)_l$ & $SO(5)$ & $U(1)_\Psi$ & $U(1)_\mathcal{S}$ \\
    \hline
    \hline
    $\Psi^3,\Psi_d^3,\mathcal{X}^{({\prime})}$ & $\mathbf{4}$ & $\mathbf{1}$ & $\mathbf{4}$ & $1$ & $0$\\
    $\Psi^j,\Psi_{u,d}^j$        & $\mathbf{1}$ & $\mathbf{4}$ & $\mathbf{4}$ & $1$ & $0$\\
    $\mathcal{S}^i$         & $\mathbf{1}$ & $\mathbf{1}$ & $\mathbf{1}$ & $0$ & $1$\\
    \hline
    \hline
    $\Sigma$ & $\mathbf{1}$ & $\mathbf{1}$ & $\mathbf{5}$ & $0$ & $0$\\
    $\Omega$ & $\mathbf{1}$ & $\mathbf{4}$ & $\mathbf{4}$ & $1$ & $-1$\\ 
    $\Phi$   & $\mathbf{1}$ & $\mathbf{1}$ & $\mathbf{1}$ & $0$ & $2$\\ 
    \hline
    \end{tabular}}
    \caption{Matter content: The upper (lower) block refers to fermions (scalars).}
    \label{tab:content}
\end{table}
The zero modes of the fermion fields correspond to the fermionic content of the 4321 model. The fermions are all in the spinorial ${\bf 4}$ representation of $SO(5)$, which decomposes as one $SU(2)_L$ doublet and one $SU(2)_R$ doublet under the IR brane symmetry. The boundary conditions for the fermions are taken to be $SU(4)$ symmetric, but $SO(5)$ breaking in the UV ($\Psi^3$ also breaks $SO(5)$ in the IR). In terms of $SU(2)_L$ doublet and $SU(2)_R$ up- and down-type components (as required by the UV brane symmetry), the boundary conditions for the fermions are
\begin{align}\label{eq:FermionBCs}
\Psi^3&=
\begin{bmatrix}
\psi^3\,(+,+)\\[2pt]
\psi_u^3\,(-,-)\\[2pt]
\tilde\psi_d^3\,(+,-)\\
\end{bmatrix}
\,,&
\Psi^3_d&=
\begin{bmatrix}
\tilde\psi^3\,(+,-)\\[2pt]
\tilde\psi_u^3\,(+,-)\\[2pt]
\psi_d^3\,(-,-)\\
\end{bmatrix}
\,, &
\mathcal{X}^{(\prime)}&=
\begin{bmatrix}
\chi^{(\prime)} (\pm,\pm)\\[2pt]
\chi^{(\prime)}_u\,(\mp,\pm)\\[2pt]
\chi^{(\prime)}_d\,(\mp,\pm)\\
\end{bmatrix}
\,,& \nonumber \\
\Psi^j&=
\begin{bmatrix}
\psi^j\,(+,+)\\[2pt]
\tilde\psi_u^j\,(-,+)\\[2pt]
\tilde\psi_d^j\,(-,+)\\
\end{bmatrix}
\,,&
\Psi_u^j&=
\begin{bmatrix}
\tilde\psi^j\,(+,-)\\[2pt]
\psi_u^j\,(-,-)\\[2pt]
\hat\psi_d^j\,(+,-)\\
\end{bmatrix}
\,,&
\Psi_d^j&=
\begin{bmatrix}
\hat\psi^j\,(+,-)\\[2pt]
\hat\psi_u^j\,(+,-)\\[2pt]
\psi_d^j\,(-,-) \\
\end{bmatrix}\,.
\end{align}
The $+$ ($-$) notation indicates that that RH (LH) chirality is zero at the given boundary, so massless zero modes only occur for $(+,+)$ (LH) or $(-,-)$ (RH) boundary conditions.
\subsection{Yukawa Couplings} Due to the gauge origin of the Higgs, the generation of Yukawa couplings requires the breaking of $SO(5)$ on both boundaries. Keeping this in mind, Yukawa couplings in the model are generated via three distinct mechanisms:
\begin{itemize}
\item {\it Top Yukawa.} The top Yukawa is special as $\Psi^3$ is the only field containing both a left ($\psi^3_L$) and right handed ($\psi^3_{uR}$) zero mode, with boundary conditions that break $SO(5)$ in both the UV and IR. This allows the top Yukawa to be generated directly in the bulk, without requiring any boundary masses. Via holographic or spectral function methods, one finds $y_t \leq g_{*} /2\sqrt{2}$, so we infer a lower bound on the $SO(5)$ KK coupling $g_{*}\geq 2\sqrt{2}\,y_t(\Lambda_{\rm IR})\approx2.2$.
\item {\it Bottom Yukawa and Light-Heavy Mixing.} The bottom and tau Yukawas, as well as the leading mixing between the light families and the fields charged under $SU(4)_h$ (such as the 3rd gen.), can only be induced after IR mixing between the 5D fields containing the relevant zero modes. The required IR masses, written in terms of $SU(2)_{L(R)}$ components in order to respect the $SO(4)$ symmetry of the IR brane, are
\begin{align}
\mathcal{L}_{\rm IR} &\supset \bar\Psi_L^3 \tilde M_{\Psi d }^L \mathcal{P}_L \Psi_{dR}^3+ \bar\Psi^j_L \tilde m_{\Psi j}^R \mathcal{P}_R \Psi_R^3 +\bar{\mathcal{X}_L}\tilde M_{\chi u}^R \mathcal{P}_R \Psi_R^3 \nonumber\\
&+ \bar\Psi^j_L (\tilde m_{dj}^L \mathcal{P}_L+\tilde m_{dj}^R \mathcal{P}_R) \Psi_{dR}^3+\bar{\mathcal{X}_L} (\tilde M_{\chi d }^L \mathcal{P}_L + \tilde M_{\chi d }^R \mathcal{P}_R)\Psi_{dR}^3 \,,
\end{align}
which yield Yukawa couplings between the zero modes of two 5D fermions $f_1$ and $f_2$ of the form
\begin{align}
y_{f1 f2}=\frac{g_*}{2\sqrt{2}}\, (\tilde M_{12}^L-\tilde M_{12}^R)\, P\big(\{c_{f_1}^{(n)}\},\{c_{f_2}^{(n)}\}\big)\,,
\end{align}
where $\tilde M_{12}^{L(R)}$ is a dimensionless IR brane mass between their $SU(2)_{L(R)}$ components and $P$ is a profile function given in Ref.~\cite{Fuentes-Martin:2022xnb}. These Yukawas vanish if $\tilde M_{12}^L=\tilde M_{12}^R$, which is the $SO(5)$-symmetric limit.
\item {\it Light-family Yukawas.} Since the RH light family fields are taken to be strongly localized in their respective branes, light-family Yukawas must be generated in the UV. This requires the introduction of an IR-localized scalar field $\Sigma$ transforming as a ${\bf 5}$ of $SO(5)$ whose singlet component takes a VEV, propagating the breaking of $SO(5)$ into the bulk. One can then write Yukawa couplings of the form 
\begin{align}
\mathcal{L}_{\rm 5D}&\supset - Y_{u,d}^{ij}\,\bar\Psi^i\,\Sigma^a\,\Gamma^a\, P_R\Psi_{u,d}^j\,,
\end{align}
leading to the generation of light-family Yukawas suppressed by the $\Sigma$ VEV profile. In the case where $\Sigma$ has a bulk mass close to the Breitenlohner-Freedman stability bound, the light family Yukawas take the following form
\begin{align}\label{eq:LightYukawas}
y_{u,d}^{ij} &\approx  \frac{g_*}{2\sqrt{2}}\, \tilde Y_{u,d}^{ij}\,
\frac{\langle \Sigma_{\rm IR}\rangle}{\Lambda_{\rm IR}}\,e^{-k (L-\ell_j)} e^{-k(c_i^{(1)}-\frac{1}{2})|y_i-\ell_j|}\,e^{k(c_j^{(1)}+\frac{1}{2})|y_j-\ell_j|}\,,
\end{align}
where the notation is explained in Ref.~\cite{Fuentes-Martin:2022xnb}. Importantly, we see that the light family Yukawas are suppressed by the hierarchy of scales $\Lambda_{\rm IR}/\Lambda_{\rm 1,2} =  e^{-k (L-\ell_{1,2})} $ as expected.
\end{itemize}
\subsection{Vector-like Fermion Mass and Mixing}
The 4321 VL fermion $\chi_{L,R}$ comes from the 5D fields $\mathcal{X}$ and $\mathcal{X}'$, whose zero modes get a mass in the IR brane. The following IR mass terms
\begin{align}
\mathcal{L}_{\rm IR}& \supset   \big(\bar{\mathcal{X}}_L \tilde M_\chi + \bar\Psi_L^3 \tilde M_{\Psi} + \bar \Psi_L^j \tilde m_\psi^j \big) \,\mathcal{P}_L \mathcal{X}^\prime_R\,,
\end{align}
generate the VL fermion mass and mix all the left-handed components of the zero modes. The vector-like masses thus induced read
\begin{align}
\mathcal{L}&\supset-M_f\,\bar f_L\chi_R^\prime\,,\;\;\hspace{10mm}
M_f=\Lambda_{\rm IR}\,  \tilde M_f\, P\big(\{c_f^{(n)}\},\{c_{\mathcal{X}^\prime}^{(n)}\}\big)\,.
\label{VLmass}
\end{align}
\section{Effective 4D Yukawa Couplings}
Applying the rules above, one can write down the effective 4D Yukawa Lagrangian in the 4321 broken phase. For the benchmark point given in~\cite{Fuentes-Martin:2022xnb}, it reads
\begin{align}
-\mathcal{L}_{\rm 4D}&\supset \frac{g_*}{2\sqrt{2}}\,\Big[\bar \psi_L^3 -\bar\chi_L \tilde M_{\chi u}^R -c_j \,  e^{-\frac{k z_j}{2}}\bar \psi_L^j \tilde m_{\Psi j}^R  \Big]\tilde H\psi_{uR}^3 + y_{u}^{ij} \bar \psi_L^i \tilde H \psi_{uR}^{j} + + y_{d}^{ij} \bar \psi_L^i H \psi_{dR}^{j}  \,  \nonumber\\ 
&+\frac{g_*}{2\sqrt{2}}\, c_2 \, e^{-\frac{k z_2}{2}}\Big[\bar\psi_L^3 \tilde M_{\Psi d}^{L} 
+\bar\chi_L(\tilde M_{\chi d }^L-\tilde M_{\chi d}^R)   +c_j \, e^{-\frac{k z_{j}}{2}} \,\bar \psi_L^j (\tilde m_{dj}^{L}-\tilde m_{dj}^{R}) \Big] H\psi_{dR}^3\nonumber \\
&+ \frac{\Lambda_{\rm IR}}{\sqrt{k L}}\,\Big[\bar\psi_L^3 \tilde M_\Psi + \bar\chi_L \tilde M_\chi + c_j \,e^{-\frac{k z_j}{2}} \,\bar\psi_L^j \tilde m_\psi^j\Big]\chi_R^\prime \, + {\rm h.c.} \,, 
\end{align}
where $i,j = 1,2$ and $z_1 = L$, $z_2 = L-\ell$, $c_1 = 1$, and $c_2 = 1/\sqrt{2}$ and there there is a suppressed index which allows the IR masses and light Yukawas to be different for quarks and leptons. This Lagrangian can then be matched to the parameters of the 4321 model, e.g. one can identify the 4321 VL fermion mass as $M_\chi = \tilde M_\chi  \Lambda_{\rm IR} / \sqrt{kL} \approx 2$ TeV. In the up sector, we find unsuppressed Yukawas, namely $y_t = g_{*}/2\sqrt{2}$ and $y_{+} = y_t \tilde M_{\chi u}^R$, while in the down sector one sees that there is an overall exponential suppression factor explaining the smallness of the bottom and tau Yukawas, coming from the localization of $\Psi_d^3$ in the 2nd family brane. There are also new sources of $U(2)_{q,\ell}$-breaking not present in the 4321 model, induced by integrating out the heavy KK partners inside the 5D fermion multiplets. Interestingly, in the near $SO(5)$ symmetric limit where all $\tilde M_L \approx \tilde M_R$, we achieve approximate down alignment for light-heavy mixing, which is phenomenologically required to pass the bounds on coloron-induced $B_{s,d} - \bar B_{s,d}$  meson mixing. An analogous limit is not possible in the up sector due to the chosen boundary conditions for $\Psi^3$.

\section{Conclusions}
We discussed a multi-scale model for flavor hierarchies based on a warped extra dimension with 3 branes where each SM family is quasi-localized. In particular, the right-handed fields are taken to be highly localized in their respective branes, resulting in a $U(2)^5$ flavor symmetry broken dominantly in the left-handed sector. This improves upon the flavor anarchy scenario, as dangerous operators involving the light-family right-handed fields are suppressed by the UV scales. The SM Higgs field emerges a pNGB from the same IR symmetry breaking that gives mass to the 4321 gauge bosons, including the $U_1$ vector leptoquark that is known to provide a good explanation of the $B$-anomalies.

\acknowledgments
I would like to thank L. Allwicher, J. M. Lizana, J. Pag\`es, and N. Selimovi{\'c} for carefully reading these proceedings as well as the organizers of La Thuile 2022 for the invitation.

\bibliographystyle{JHEP}
\bibliography{stefanek_proceedings.bib}

\providecommand{\href}[2]{#2}\begingroup\raggedright\begin{thebibliography}{10}

\bibitem{Fuentes-Martin:2022xnb}
J.~Fuentes-Martin, G.~Isidori, J.~M. Lizana, N.~Selimovic, and B.~A. Stefanek,
  {\it {Flavor hierarchies, flavor anomalies, and Higgs mass from a warped
  extra dimension}},  \href{http://arxiv.org/abs/2203.0195}{{\tt
  arXiv:2203.0195}}.

\bibitem{Randall:1999ee}
L.~Randall and R.~Sundrum, {\it {A Large mass hierarchy from a small extra
  dimension}},  {\em Phys. Rev. Lett.} {\bf 83} (1999) 3370--3373,
  [\href{http://arxiv.org/abs/hep-ph/9905221}{{\tt hep-ph/9905221}}].

\bibitem{DiLuzio:2017vat}
L.~Di~Luzio, A.~Greljo, and M.~Nardecchia, {\it {Gauge leptoquark as the origin
  of B-physics anomalies}},  {\em Phys. Rev. D} {\bf 96} (2017), no.~11 115011,
  [\href{http://arxiv.org/abs/1708.0845}{{\tt arXiv:1708.0845}}].

\bibitem{Bordone:2017bld}
M.~Bordone, C.~Cornella, J.~Fuentes-Martin, and G.~Isidori, {\it {A three-site
  gauge model for flavor hierarchies and flavor anomalies}},  {\em Phys. Lett.
  B} {\bf 779} (2018) 317--323, [\href{http://arxiv.org/abs/1712.0136}{{\tt
  arXiv:1712.0136}}].

\bibitem{Greljo:2018tuh}
A.~Greljo and B.~A. Stefanek, {\it {Third family quark\textendash{}lepton
  unification at the TeV scale}},  {\em Phys. Lett. B} {\bf 782} (2018)
  131--138, [\href{http://arxiv.org/abs/1802.0427}{{\tt arXiv:1802.0427}}].

\bibitem{Fuentes-Martin:2020bnh}
J.~Fuentes-Mart\'\i{}n and P.~Stangl, {\it {Third-family quark-lepton
  unification with a fundamental composite Higgs}},  {\em Phys. Lett. B} {\bf
  811} (2020) 135953, [\href{http://arxiv.org/abs/2004.1137}{{\tt
  arXiv:2004.1137}}].

\bibitem{Alonso:2015sja}
R.~Alonso, B.~Grinstein, and J.~Martin~Camalich, {\it {Lepton universality
  violation and lepton flavor conservation in $B$-meson decays}},  {\em JHEP}
  {\bf 10} (2015) 184, [\href{http://arxiv.org/abs/1505.0516}{{\tt
  arXiv:1505.0516}}].

\bibitem{Calibbi:2015kma}
L.~Calibbi, A.~Crivellin, and T.~Ota, {\it {Effective Field Theory Approach to
  $b\to s\ell\ell^{(')}$, $B\to K^{(*)}\nu\overline{\nu}$ and $B\to
  D^{(*)}\tau\nu$ with Third Generation Couplings}},  {\em Phys. Rev. Lett.}
  {\bf 115} (2015) 181801, [\href{http://arxiv.org/abs/1506.0266}{{\tt
  arXiv:1506.0266}}].

\bibitem{Barbieri:2015yvd}
R.~Barbieri, G.~Isidori, A.~Pattori, and F.~Senia, {\it {Anomalies in
  $B$-decays and $U(2)$ flavour symmetry}},  {\em Eur. Phys. J. C} {\bf 76}
  (2016), no.~2 67, [\href{http://arxiv.org/abs/1512.0156}{{\tt
  arXiv:1512.0156}}].

\bibitem{Bhattacharya:2016mcc}
B.~Bhattacharya, A.~Datta, J.-P. Gu\'evin, D.~London, and R.~Watanabe, {\it
  {Simultaneous Explanation of the $R_K$ and $R_{D^{(*)}}$ Puzzles: a Model
  Analysis}},  {\em JHEP} {\bf 01} (2017) 015,
  [\href{http://arxiv.org/abs/1609.0907}{{\tt arXiv:1609.0907}}].

\bibitem{Buttazzo:2017ixm}
D.~Buttazzo, A.~Greljo, G.~Isidori, and D.~Marzocca, {\it {B-physics anomalies:
  a guide to combined explanations}},  {\em JHEP} {\bf 11} (2017) 044,
  [\href{http://arxiv.org/abs/1706.0780}{{\tt arXiv:1706.0780}}].

\bibitem{DiLuzio:2018zxy}
L.~Di~Luzio, J.~Fuentes-Martin, A.~Greljo, M.~Nardecchia, and S.~Renner, {\it
  {Maximal Flavour Violation: a Cabibbo mechanism for leptoquarks}},  {\em
  JHEP} {\bf 11} (2018) 081, [\href{http://arxiv.org/abs/1808.0094}{{\tt
  arXiv:1808.0094}}].

\bibitem{Cornella:2019hct}
C.~Cornella, J.~Fuentes-Martin, and G.~Isidori, {\it {Revisiting the vector
  leptoquark explanation of the B-physics anomalies}},  {\em JHEP} {\bf 07}
  (2019) 168, [\href{http://arxiv.org/abs/1903.1151}{{\tt arXiv:1903.1151}}].

\bibitem{Cornella:2021sby}
C.~Cornella, D.~A. Faroughy, J.~Fuentes-Martin, G.~Isidori, and M.~Neubert,
  {\it {Reading the footprints of the B-meson flavor anomalies}},  {\em JHEP}
  {\bf 08} (2021) 050, [\href{http://arxiv.org/abs/2103.1655}{{\tt
  arXiv:2103.1655}}].

\bibitem{Panico:2016ull}
G.~Panico and A.~Pomarol, {\it {Flavor hierarchies from dynamical scales}},
  {\em JHEP} {\bf 07} (2016) 097, [\href{http://arxiv.org/abs/1603.0660}{{\tt
  arXiv:1603.0660}}].

\bibitem{Contino:2003ve}
R.~Contino, Y.~Nomura, and A.~Pomarol, {\it {Higgs as a holographic
  pseudoGoldstone boson}},  {\em Nucl. Phys. B} {\bf 671} (2003) 148--174,
  [\href{http://arxiv.org/abs/hep-ph/0306259}{{\tt hep-ph/0306259}}].

\bibitem{Agashe:2004rs}
K.~Agashe, R.~Contino, and A.~Pomarol, {\it {The Minimal composite Higgs
  model}},  {\em Nucl. Phys. B} {\bf 719} (2005) 165--187,
  [\href{http://arxiv.org/abs/hep-ph/0412089}{{\tt hep-ph/0412089}}].

\bibitem{Greljo:2019xan}
A.~Greljo, T.~Opferkuch, and B.~A. Stefanek, {\it {Gravitational Imprints of
  Flavor Hierarchies}},  {\em Phys. Rev. Lett.} {\bf 124} (2020), no.~17
  171802, [\href{http://arxiv.org/abs/1910.0201}{{\tt arXiv:1910.0201}}].

\bibitem{Fuentes-Martin:2020pww}
J.~Fuentes-Martin, G.~Isidori, J.~Pag\`es, and B.~A. Stefanek, {\it {Flavor
  non-universal Pati-Salam unification and neutrino masses}},  {\em Phys. Lett.
  B} {\bf 820} (2021) 136484, [\href{http://arxiv.org/abs/2012.1049}{{\tt
  arXiv:2012.1049}}].

\end{thebibliography}\endgroup

\end{document}